\documentclass[11pt,a4paper,onecolumn,reqno]{amsart}
\usepackage[a4paper, margin=2.3cm, bmargin=3cm]{geometry}

\usepackage{graphicx,stfloats}
\usepackage{epstopdf}
\usepackage{color}
\usepackage{amsmath}
\usepackage{amsaddr}
\usepackage{bm, soul}
\usepackage{mathtools}
\usepackage[colorinlistoftodos,prependcaption,textsize=tiny]{todonotes}
\usepackage{braket}
\usepackage{hyperref}
\usepackage{makecell}
\usepackage{array}
\usepackage{placeins}
\usepackage[square,comma,sort&compress, numbers]{natbib}

\usepackage[version=3]{mhchem} 

 \usepackage{multirow}

\usepackage{etoolbox}
\patchcmd{\pprintMaketitle}
  {\hrule\vskip12pt}
 {\hrule\vskip12pt\ifvoid\extrainfobox\else\unvbox\extrainfobox\par\vskip12pt\fi}
 {}{}

\newsavebox\extrainfobox

\usepackage{caption}
\usepackage[utf8]{inputenc}
\usepackage{amssymb}
\usepackage{amsmath}
\usepackage{hyperref}
\usepackage{ulem}
\usepackage[export]{adjustbox}
\usepackage{xcolor}
\usepackage{nicefrac}
\usepackage{tabu,multirow}
\usepackage{tabularx}
\def\hlinewd#1{%
\noalign{\ifnum0=`}\fi\hrule \@height #1 %
\futurelet\reserved@a\@xhline}
\usepackage{graphicx}
\usepackage{amssymb}

\usepackage{lineno}
\usepackage{color}
\usepackage{float}
\usepackage{epstopdf}
\usepackage{siunitx}
\usepackage{hyphenat}
\usepackage{todonotes}
\usepackage{subcaption}
\usepackage[version=3]{mhchem} 
\usepackage{gensymb}

\DeclareSIUnit\volper{vol\%}
\DeclareSIUnit\massper{\%_{m}}
\usepackage{array}
\DeclareMathAlphabet{\mathpzc}{OT1}{pzc}{m}{it}
\newcolumntype{P}[1]{>{\centering\arraybackslash}p{#1}}
\hypersetup{
colorlinks=true,
linkcolor=blue,
citecolor=blue,
citebordercolor={0 1 0}
}

\title{Iron as a sustainable chemical carrier of renewable energy: Analysis of opportunities and challenges for retrofitting coal-fired power plants}

\author[stfs]{Paulo Debiagi$^{1,*}$, Rodolfo Cavaliere da Rocha$^1$, Arne Scholtissek$^1$, Johannes Janicka$^2$, Christian Hasse$^1$}
\email{debiagi@stfs.tu-darmstadt.de} 

\address[]{$^1$ Technical University of Darmstadt, Department of Mechanical Engineering, Simulation of reactive Thermo-Fluid Systems, Otto-Berndt-Str. 2, 64287 Darmstadt, Germany \\
$^2$Technical University of Darmstadt, Department of Mechanical Engineering, Energy and Power Plant Technology, Otto-Berndt-Straße 3, 64287 Darmstadt, Germany}

\begin{document}
\pagestyle{plain}
\maketitle

\begin{abstract}
As a result of the 2021 United Nations Climate Change Conference (COP26), several countries committed to phasing down coal electricity as soon as possible, deactivating hundreds of power plants in the near future. \ce{CO2}-free electricity can be generated in these plants by retrofitting them for iron combustion. Iron oxides produced during the process can be collected and reduced back to metallic iron using \ce{H2}, in a circular process where it becomes an energy carrier. Using clean energy in the recycling process enables storage and distribution of excess generated in periods of abundance. This concept uses and scales up existing dry metal cycle technologies, which are the focus of extensive research worldwide. Retrofitting is evaluated here to determine feasibility of adding these material requirements to markets, in the context of current plans for decarbonization of steel industry, and policies on hydrogen and renewable electricity. Results indicate that not only for a single power plant, but also on larger scales, the retrofitting plan is viable, promoting and supporting advancements in sustainable electricity, steel industry and hydrogen production, converging necessary technological and construction efforts. The maturation and first commercial-scale application of iron combustion technology by 2030, together with developing necessary reduction infrastructure over the next decades, would pave the way for large-scale retrofitting and support the phasing out of coal in many regions. The proposed plan represents a feasible solution that takes advantage of existing assets, creates a long-lasting legacy for the industry and establishes circular energy economies that increase local energy security.\\

\end{abstract}

\keywords{\textbf{Keywords:} Iron Combustion; Energy Carrier; Retrofitting; Green Energy; Hydrogen; Sustainability; Carbon-free Combustion}
\clearpage

\section{Introduction}
\begin{figure}[h]
        \begin{center}
        \includegraphics[width=1\textwidth]{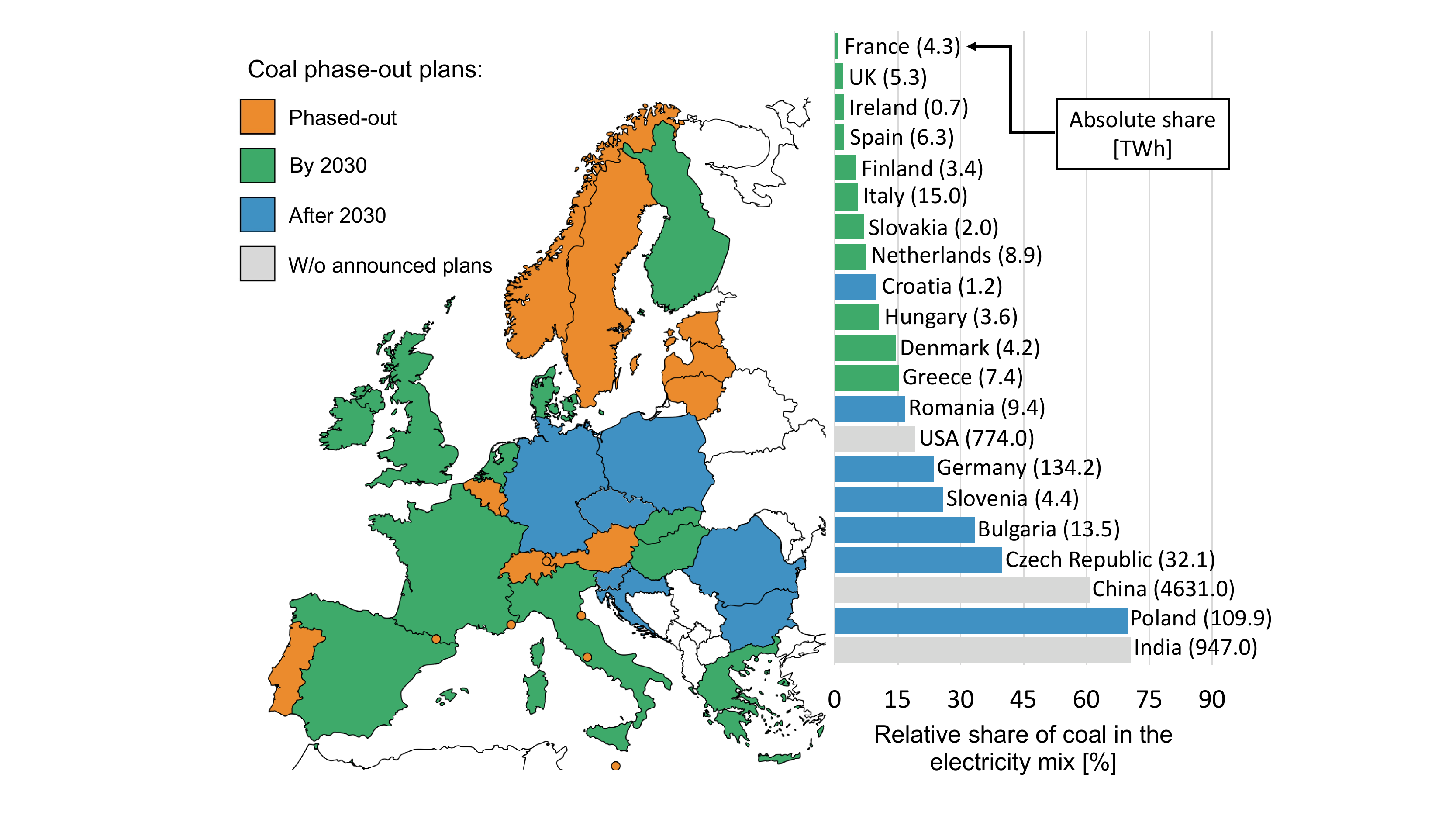}
        \end{center}
        \caption{\textbf{Share of coal in electricity mix and plans for phasing out coal in Europe.} Left: Current plans for phasing out coal in selected European countries. Right: Absolute [TWh] and relative [\%] share of coal in the electricity mix of selected European countries (2020) \cite{emberWorld,emberEurope}, clearly indicating that countries that are less reliant on coal are committing to phase out coal sooner. The share in China, India and the USA is also shown. \cite{coalexittracker, poland2049}.}
		\label{fig:Europe_phase-out}
\end{figure}

In the Paris Agreement, countries committed to reducing emissions of greenhouse gases (GHGs), limiting global warming, preferably to 1.5 $^\circ$C. Investigations by the Intergovernmental Panel on Climate Change (IPCC) forecast that the average global temperature would reach 1.4--4.4 $^\circ$C above pre-industrial times by 2100 \cite{AR6full2021}. During the 2021 United Nations Climate Change Conference (COP26), the parties agreed to phase down unabated coal usage \cite{COP26}, which was the source of about 30\% of total global \ce{CO2} emissions in 2020 \cite{IEAwebsite,BP2021}. Coal combustion is also the source of pollutants such as heavy metals, particulate matter, nitrogen oxides (NO$_x$) \cite{zhang2019effect} and sulfur oxides (SO$_x$) \cite{cai2021novel}. The global electricity sector relied on coal for 35.1\% (9.4 PWh) of the output in 2020 \cite{BP2021}, a decrease compared with the historical 2018 peak (10.1 PWh), partly because of the COVID-19 pandemic. Percentages vary locally, at 60.75\% in China \cite{emberWorld}, 23.66\% in Germany and 13.22\% in EU-27 \cite{emberEurope}. Currently (2021), 2445 coal power stations are operating around the world, totalling 2.07 TW of capacity \cite{endcoal2021}. Several European countries have established plans to phase out coal by 2030, as reported in Fig.~\ref{fig:Europe_phase-out}, clearly showing that the treansition is easier for all these countries, which require less than 15\% of coal in their electricity mix. Germany previously announced plans to phase out coal by 2038 \cite{coalexittracker}, and the recently formed government proposes bringing this forward to 2030 \cite{coalitionDE2021}. Poland is a special case in the EU, as it still relies strongly on coal (69.8\%) \cite{emberEurope}, making this transition very challenging. Further resistance to phasing coal out comes from India and China, which are large consumers of electricity in general, and are particularly reliant on coal \cite{emberWorld}.  

Compared to other fossil fuels, in 2020 coal is still the largest source of global energy-related \ce{CO2} emissions (44.0\%), followed by oil and its derivates (33.7\%), and natural gas (21.6\%). Many countries are investing in power generation from natural gas to support the phasing out of coal, as both pollutants and \ce{CO2} emissions are significantly lower. However, the current crisis in Ukraine is expected to cause rapid policy changes worldwide, especially in Europe, where several sectors are pressuring policy-makers to reduce and eventually cease energy imports from Russia. Sustainable processes for converting renewable energy sources need to be developed \cite{sher2021sustainable} and this will depend on multiple solutions in the future, including e-fuels \cite{al2020effect,al2020electrochemical}, biofuels \cite{yusoff2021solvent,rashid2021effect} and carbon capture \cite{qureshi2021part,sher2020development}. Renewable energy sources (e.g., wind, sun, geothermal, biomass) are generally proposed as substitutes for fossil fuels for electricity generation. In recent years, governments worldwide have been implementing policies to enable a fast transition of the electricity mix, scaling up the capacity. The new German coalition presented plans to ramp up the clean energy output from 251 TWh in 2020 \cite{RenewDE2021} to as much as 600 TWh by 2030 (80\% of the power mix), later increasing it to 100\% of the total output by 2045, when the country is expected to become carbon neutral \cite{coalitionDE2021}. Once deactivated, existing infrastructure for fossil fuels can remain as a backup capacity, or undergo retrofitting processes for renewable energy. These initiatives, along with technology maturation, have led to a steep decline in renewable electricity costs, especially for solar and wind, to the point where it has become considerably cheaper than fossil fuel-based electricity in places where resources are abundant \cite{WEO2020,Luderer2021,IRENA2020}.

\begin{figure}[h]
		\begin{center}
        \includegraphics[width=0.9\textwidth]{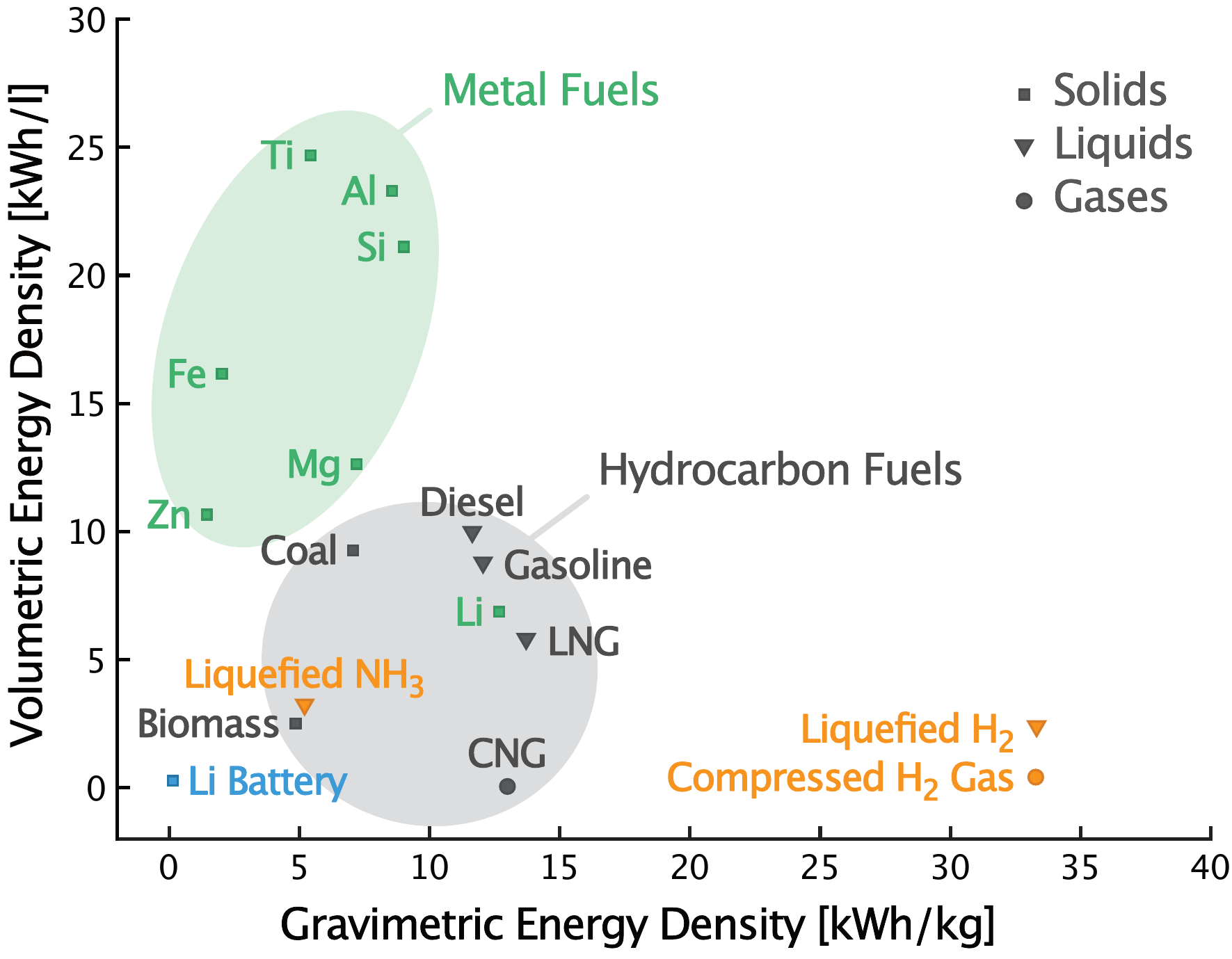}
        \end{center}
        \caption{\textbf{Volumetric and gravimetric energy densities of chemical energy carriers}.  Representative hydrocarbon fuels are characterized by a balance between the two properties. \ce{H2} has a high gravimetric energy density but a low volumetric density, even when compressed or liquefied. In contrast, metals have a very high volumetric energy density with a low to medium gravimetric energy density. Adapted from Bergthorson \cite{bergthorson2018}.}
		\label{fig:EnergyDensity}
\end{figure}

Despite this progress, a complete transition to alternative energy sources remains very challenging. Renewable resources are not available everywhere, meaning that the viability of power plants and their relative production costs are related to their geographic location. This also leads to a distribution problem, since the main suppliers and the main consumers may be very distant from each other. Additionally, the sun and wind provide an intermittent and unpredictable supply of power, and the excess energy produced in periods of abundance must be stored for use in periods of scarcity \cite{dreizler_2021,AYODELE2015447}. Energy storage and distribution are a challenge and require the use of cost-effective energy carriers~\cite{bergthorson2018}. Due to their high energy density and versatility, chemical energy carriers are suggested for long-distance energy trading, remote power generation, heavy-duty machinery and transportation equipment \cite{bergthorson2018}. Several candidates have been proposed and some are currently being investigated as eligible solutions, including hydrogen \cite{thefutureofhydrogen}, ammonia \cite{valera-medina2021}, and metals \cite{BERGTHORSON2015368}, as reported in Fig. \ref{fig:EnergyDensity}. 

In general, metals have long been used as energy carriers to store and release electrical energy, which is the main working mechanism of commercial batteries \cite{Yang2021}. They are also frequently employed to produce hydrogen on demand, through their reaction with water or basic solutions \cite{julien2017,Niether2018}. Metal-fuel cycles, in which metals are employed as recyclable zero-carbon fuels for large-scale power generation, have recently been suggested. Bergthorson et al. \cite{bergthorson2018,BERGTHORSON2015368} extensively analyse the dry metal-fuel cycles that can be efficiently employed for zero-carbon power generation. They also review the wet metal-fuel cycle using water for clean propulsion and power generation \cite{BERGTHORSON201713}. Shkolnikov et al. \cite{SHKOLNIKOV20114611} provides an overview of the feasibility of aluminum as a metal fuel to generate power, and the technology required. Dirven et al. \cite{DIRVEN201852} provide an assessment and compare coal with four metal fuels in terms of their overall efficiency on a complete cycle of power generation and fuel regeneration. Julien et al. \cite{julien2017} comprehensively evaluate different metals based on five constraints—reactivity with oxygen (1), competitive energy density (2), toxicity, radioactivity and safety (3), reduction without \ce{CO2} emissions (4) and scalability of the technology (5)—and identify boron, magnesium, aluminum, silicon, titanium and iron as feasible energy carriers. As stated by the authors \cite{julien2017}, metal fuels can be used, captured, and cyclically reduced back into metal fuels an indefinite number of times. 

Clean Circles cluster project \cite{cleancircles} teams up scientists from multiple disciplines to explore how metals and their oxides can be used in a cycle as carbon-free chemical energy carriers to store wind and solar power. Several research groups have been working on developing dry iron-fuel technologies. Goroshin et al. \cite{GOROSHIN2011656} performed reduced-gravity experiments on laminar iron dust flames and Tang et al. \cite{TANG20111975} modeled the modes of combustion for iron dust particles using those experimental data. Hazenberg and van Oijen \cite{hazenberg2021} developed an Eulerian-Lagrangian model to capture the flame structure and burning velocities of iron dust flames. Huang et al. \cite{huang2021} performed experiments to improve the understanding and prediction of micro-explosion events in pulverized iron combustion. Ning et al. \cite{ning2022} experimentally measured the time-resolved temperature for iron particles using laser ignition, and characterised several properties of the post-combustion particles.

\begin{figure}[h]
		\begin{center}
        \includegraphics[width=0.9\textwidth]{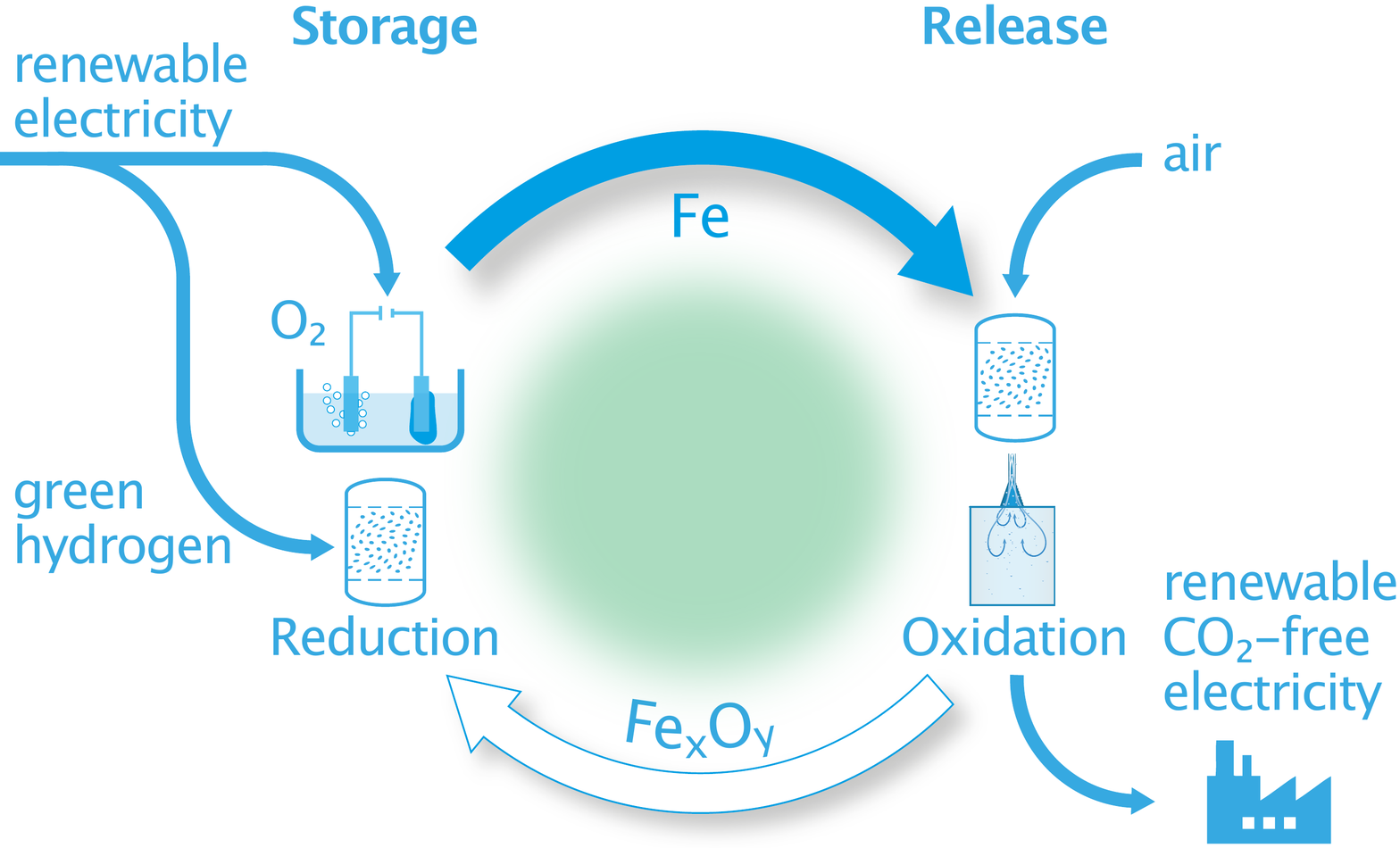}
        \end{center}
        \caption{\textbf{Schematic of an iron reduction-oxidation cycle for a \ce{CO2}-free energy supply.} Adapted from \cite{bergthorson2018}. Iron and iron oxides are used in a reduction-oxidation cycle as carbon-free carriers of renewable energy. On the right-hand side, electricity is generated (Release) using iron as a fuel that is burned to release heat during high-temperature oxidation, similarly to the traditional combustion of solid fuels. Iron combustion does not produce \ce{CO2} emissions. Solid iron oxides (Fe$_x$O$_y$) are the  combustion products, which can easily be captured. In the bottom part of the cycle, iron oxides are  stored and transported to the reduction facilities. Renewable energy is used to chemically reduce iron oxides via electrochemical or thermochemical processes (Storage). Green \ce{H2} is used as a reducing agent for the thermochemical route. Both options regenerate iron fuel for the combustion process without \ce{CO2} emissions. The cycle is closed when recycled iron fuel is transported to the power plant (upper part), establishing a circular energy economy.}
		\label{fig:MFcycle}
\end{figure}

Iron, in particular, is very attractive for stationary power generation, and is suggested in the context of a green metal fuel economy \cite{bergthorson2018}. It is stable and abundant, has a high volumetric energy density, low toxicity and low market price, and it boasts an existing production infrastructure and transportation network \cite{bergthorson2018}. It can be burned to produce heat during high-temperature oxidation, similarly to the traditional combustion of solid fuels, but without releasing \ce{CO2}. The products of its combustion are iron oxides, which are solid under standard conditions and can be easily collected and recycled. A clean, dry oxidation/reduction cycle for using iron as an energy carrier is depicted in Fig. \ref{fig:MFcycle}. Due to its fuel characteristics, iron can in theory be used instead of coal in existing power plants, requiring only limited modifications for retrofitting, including changes related to higher particle loading, air/particle separation and post-combustion particle collection \cite{BERGTHORSON2015368}. This could pave the way for using present-day infrastructure and rapidly replacing coal. In a circular economy concept (e.g, Fig. \ref{fig:MFcycle}), clean primary energy sources are combined with innovative iron reduction techniques \cite{julien2017,auner2006,bardsley2008,wen2010,matsui2013} that use hydrogen as an intermediate \cite{MIGNARD20075039,LORENTE20095554} to recycle the metal numerous times. In fact, the quest to decarbonize the iron and steel industry, one of the most \ce{CO2}-intensive activities, goes hand in hand with this idea, the scaling up of such production processes being extremely beneficial to the industry in the long-term \cite{ISTR2020}. Siderwin \cite{siderwin}, ArcelorMittal \cite{ArcelorMittal} and ThyssenKrupp \cite{ArcelorMittal} are examples of ongoing large projects in this direction.

Completely retrofitting the world's active coal power plants for iron combustion would save approximately 10 Mt of \ce{CO2} emissions a year \cite{IEAwebsite}. However, shutting down coal power plants is a challenge, mainly for countries that rely on generating a large share of their electricity from coal. Developing solutions that could keep power plants active using existing equipment for low-emission combustion has enormous potential.
This work examines the materials and infrastructure required for retrofitting coal power plants using iron as a metal fuel, presenting a multi-scale feasibility evaluation for implementing this concept, supported by current and forecast data on iron, hydrogen and renewable energy production. First, the retrofitting scheme is analyzed at the scale of a large power station in Germany, compared to German steel production in terms of the availability of iron on the market, and to current plans for renewable energy and \ce{H2} production, which are essential for the recycling process. The analysis is then expanded to the European and global scales, evaluating the concept's feasibility when implemented in long-term plans, which involve building up inventories of iron fuel at the same time as expanding the iron reduction infrastructure and the maturation of the technology. The benefits and challenges of this concept are discussed throughout the manuscript.

\section{Overview of the dry iron redox cycle}
\label{sec:Iron_Redox_cycle}

In metallic form, iron reacts with oxygen to generate two main stable oxides: magnetite (\ce{Fe3O4}) and hematite (\ce{Fe2O3}). The process follows the global reactions:
\begin{equation}
    3\cdot \ce{Fe} + 2 \cdot \ce{O2} \rightarrow \ce{Fe_3O_4} \qquad \Delta H_{\mathrm{r}}^\circ (298 \textrm{K})= -1118.3\textrm{ kJ.mol}^{-1}
    \label{eq:Fe_oxidation_Fe3O4}
\end{equation}
\begin{equation}
    2\cdot \ce{Fe} + \frac{3}{2} \cdot \ce{O2} \rightarrow \ce{Fe_2O_3} \qquad \Delta H_{\mathrm{r}}^\circ (298 \textrm{K})= -824.2 \textrm{ kJ.mol}^{-1}
    \label{eq:Fe_oxidation_Fe2O3}
\end{equation}

These oxides are non-toxic, and are present in several minerals and compounds existing in nature. The global reactions are exothermic and, at high temperatures the thermochemical oxidation occurs rapidly, releasing usable energy that can be harnessed in thermal energy systems.

When burned in laminar or turbulent flames formed from the suspension powders or sprays of micron-sized particles, iron has combustion characteristics close to those of hydrocarbon fuels in terms of heat release, flame temperature and burning velocity \cite{BERGTHORSON2015368,TANG20111975,GOROSHIN2011656}. Iron combustion is a heterogeneous process, in which oxygen reacts on the particle surface, forming a solid oxide layer, which controls the reaction progress by diffusion. This process, in stoichiometric to fuel-rich conditions, would generate no gaseous products apart from heated \ce{N2}, while producing oxide-metal particles. These are slightly heavier and larger than the original particles used as fuel, making them simple to collect and recycle later \cite{BERGTHORSON2015368}. From the point of view of safety, the absence of nanoparticles in the products is also an advantage when compared to other metals, since these can be emitted as aerosols in the exhaust system, or interact with parts of the burner \cite{bergthorson2018}.

Systems where pulverized solid fuel clouds are burned, such as those in modern coal and biomass power plants, present technologies similar to those necessary for iron combustion, in terms of injection and energy harnessing techniques. Differently from these fuels, pollutant formation from iron combustion is minimal: it does not contain fuel-bound nitrogen and sulfur, which lead to the formation of nitrogen and sulphur oxides (NO$_x$ and SO$_x$, respectively), and it does not produce unburned hydrocarbons, which lead to formation of particulate matter (soot). However, since temperatures above 1200 $^\circ$C promote the oxidation of the nitrogen found in the combustion air (thermal-NO$_x$ formation path) \cite{zhang2019effect}, NO$_x$ formation may have to be controlled by maintaining existing techniques: deep-air staging, reburning and selective catalytic and non-catalytic NO$_x$ reduction \cite{Niksa2019}. Additionally, ash collection devices used for conventional solid fuels could be adapted to collect the solid oxides produced in the combustion, while the heated \ce{N2} could be used for subsequent thermal and chemical processes.

Iron oxides can be recycled back into metallic iron using a reducing agent. Taking conventional steel production using blast furnaces, the iron oxides present in iron ore are reduced by the carbon monoxide released by the partial oxidation of coke, forming \ce{CO2} in a very intensive GHG-emitting process. One promising alternative is to use hydrogen instead of coke, yielding \ce{H2O} instead of \ce{CO2} as the by-product. The positive environmental impact is clear, as anthropogenic \ce{H2O} vapor emissions have negligible climate warming effects \cite{Sherwood_2018}. The global reduction reactions of the stablest iron oxides can be described globally as:
\begin{equation}   \ce{Fe_3O_4}+ 4\cdot \ce{CO} \rightarrow 3\cdot \ce{Fe} +  4\cdot \ce{CO2} \qquad \Delta H_{\mathrm{r}}^\circ (298 \textrm{K})= -13.6 \textrm{ kJ.mol}^{-1}
   \label{eq:C_reduction_1}
\end{equation}
\begin{equation}
   \ce{Fe_2O_3} + 3\cdot \ce{CO} \rightarrow 2\cdot \ce{Fe} + 3\cdot \ce{CO2} \qquad \Delta H_{\mathrm{r}}^\circ (298 \textrm{K})= -24.7 \textrm{ kJ.mol}^{-1}
   \label{eq:C_reduction_2}
\end{equation}
\begin{equation}
   \ce{Fe_3O_4} + 4 \cdot \ce{H2} \rightarrow 3\cdot \ce{Fe} + 4 \cdot \ce{H2O} \qquad \Delta H_{\mathrm{r}}^\circ (298 \textrm{K})= +151.0 \textrm{ kJ.mol}^{-1}
      \label{eq:H2_reduction_1}
\end{equation}
\begin{equation}
   \ce{Fe_2O_3} + 3 \cdot \ce{H2} \rightarrow 2\cdot \ce{Fe} + 3 \cdot \ce{H2O} \qquad \Delta H_{\mathrm{r}}^\circ (298 \textrm{K})= +98.76 \textrm{ kJ.mol}^{-1}
      \label{eq:H2_reduction_2}
\end{equation}

Once metallic iron is regenerated by \ce{H2} reduction, it can be used again as a fuel in thermal power plants, closing the redox cycle.

\section{Relevant properties of coal and iron as solid fuels}
\label{sec:fuel_properties}
Estimating the materials required to replace coal with iron in power plants requires the comparison of these fuels' properties, as reported in Table \ref{table:fuel_properties}. Bituminous, sub-bituminous and lignite coal types are typically used in power plants, and they mostly differ in terms of their energy density. Iron and iron oxides are highly dense materials. The iron oxidation process produces iron oxides as by-products. As observed, the heating value in the oxides decreases progressively in line with the oxidation state, reaching zero in the hematite form (\ce{Fe2O3}). Compared with typical bituminous coals, iron exhibits a much higher specific mass (density), but a significantly lower heating value the basis of mass. However, the volumetric energy density of iron is significantly higher, which is advantageous, since the volume occupied in transportation and storage units and the volumetric flow in burners would be smaller than those of coal for the same energy content. 

\begin{table}[h]\small
\caption{Properties of iron, iron oxides and different coal ranks \cite{nist,LETT2004411}.}
\centering
\begin{tabular}{l c c c c}
    \hline
    Fuel & Density & Melting Point & LHV & LHV  \\
     & [kg m$^{-3}$] & [$^\circ$C] & [MJ kg$^{-1}$] & [GJ m$^{-3}$]\\
    \hline
    \ce{Fe}     & 7870      & 1538  & 7.36      & 57.92 \\
    \ce{FeO}    & 5740      & 1377  & 2.03      & 11.65 \\
    \ce{Fe3O4}  & 5180      & 1590  & 0.35      & 1.81 \\
    \ce{Fe2O3}  & 5260      & 1565  & 0         & 0 \\
    Bituminous Coal   & 1200--1600  & - & 25--36    & 30--57  \\
    Sub-Bit. Coal    & 900--1200     & - & 20--25    & 18--30  \\
    Lignite     & 700--900      & - & 10--20    & 7--18\\
    \hline
\end{tabular}
\label{table:fuel_properties}
\end{table}

\section{Review of the current situation and plans for iron, steel and hydrogen}
\label{sec:Review_iron_h2}

This section provides an assessment of current and future plans for the two key material resources for the proposed technology, iron and hydrogen, to measure the feasibility of the retrofit program's requirements. The present-day challenges and opportunities in the iron and steel industry are evaluated, providing a roadmap for the carbon-neutral industrial production of such materials, focusing on processes that use \ce{H2} as a carbon-free reduction agent and an energy carrier. Further, an analysis of the situation regarding hydrogen is conducted in light of the current plans of the industry and governmental bodies to ramp up clean production in the broader context of what is called the hydrogen economy \cite{thefutureofhydrogen}.

\subsection{Scenario for the present and future of the iron and steel industry}

The iron and steel industry has great potential to enable the production of iron for use as an energy carrier. Around 1.9 Gt of reduced iron/steel are produced yearly, being the third most abundant man-made bulk material \cite{ISTR2020}, surpassed only by the production of sand/gravel (40--50 Gt) and cement (4.2 Gt) \cite{MineralCommoditySummaries2022}. In 2020, iron had a higher production output than any other metal, surpassing aluminum (65.1 Mt), copper (25.3 Mt) and zinc (12 Mt) by several orders of magnitude \cite{MineralCommoditySummaries2022}. The material is widely used for a number of applications in sectors such as construction, transportation, machinery and consumer goods. Production processes are well-established and technologically diversified, and, despite the current concentration in a few countries, iron has the potential to be produced worldwide.

Presently, steel production is one of the most energy-intensive industries in the world. In 2019 alone, it accounted for 9.83 PWh of energy consumption. That is the equivalent to 20\% of the total industrial energy use and 8\% of the total final energy use \cite{ISTR2020,QUADER2015594}. Almost 75\% of this energy comes from coal. Adding to that number, it is estimated that 16\% of the coal consumed globally in a year is used as coke (around 872 Mt), a reduction agent and an alloy solute. Apart from that, around 90 bcm (billion cubic meters) of natural gas, and 1.23 PWh of electricity, are used by the steel industry annually worldwide, comprising 2.5\% and 5.5\% of the global consumption of these resources, respectively. In a whole year, 2.6 Gt of \ce{CO2} are emitted from the process.

Several production processes have different demands and impacts in terms of \ce{CO2} emissions and energy consumption, see Table \ref{table:iron_production}. Primary steel, i.e., steel produced using iron ore as the primary metallic input, requires three main steps: raw material preparation, ironmaking through reduction, and steelmaking. The most common production pathway, the blast-furnace-basic oxygen furnace (BF-BOF), removes oxygen from prepared pellets or sinter by applying reduction processes in a liquid state on a furnace, using coal or natural gas for chemical reactions and for heat generation, followed by carbon removal in a second furnace by reacting the resulting ``pig iron" with limestone, releasing it in the form of \ce{CO2}. It is a mature technology that can use iron ore of variable quality. The entire process requires considerable amounts of energy, and has high indirect \ce{CO2} emissions due to electricity and heat consumption. The DRI-EAF pathway is an alternative route mostly followed on smaller scales, consisting in the direct reduction of iron (DRI), generally reacting syngas—i.e., a mixture of \ce{CO} and \ce{H2}, conventionally produced from natural gas—with solid pellets in a reactor to remove the oxygen content, followed by steelmaking in electric arc furnaces (EAFs). It is a relatively mature technology that has the disadvantage of requiring high-quality iron ore. It calls for slightly less energy and \ce{CO2} than BF-BOF, but has fewer indirect emissions, mainly related to power generation. \cite{ISTR2020}.

Secondary steel, i.e., steel produced using scrap as the primary metallic input, is the least energy- and emission-intensive process, and requires the melting and processing of residual steel in electric furnaces, especially electric arc furnaces, which consume mostly electricity and very small amounts of fossil fuels. This process only requires, on average, 10\% of the energy of BF-BOF, with much fewer direct and indirect emissions, the latter only being associated with the electricity generation \cite{ISTR2020,HARVEY2021110553}.

\begin{table}[h!]
\caption{Overview of energy and emissions impact of steel production \cite{ISTR2020}.}
\centering
\begin{tabular}{c  c  c c  c}
    \hline
    Production process  & Energy intensity & \multicolumn{3}{c}{\ce{CO2} emissions [t \ce{CO2}/t steel]}\\
     \cline{3-5} & [MWh/t steel] & Direct & Indirect & Total \\
    \hline
    BF-BOF & 5.94 & 1.2 & 1.0 & 2.2  \\
    DRI-EAF & 4.75 & 1.0 & 0.4 & 1.4\\
    Secondary & 0.58 & 0.04 & 0.3 & 0.34\\
    \hline
\end{tabular}
\label{table:iron_production}
\end{table}

\subsubsection{Roadmap for green steel production}
\label{sec:Roadmap_green_steel}
Presently, about 80\% of production involves primary steel (1.51 Gt/year), while only 20\% (0.38 Gt/year) comes from secondary steel \cite{ISTR2020}. Global demand is expected to increase by 0.6 Gt by 2050 \cite{ISTR2020}. In a sustainable development scenario (i.e., where efforts are made to improve production processes and optimize material usage), the demand will be roughly the same as today, but \ce{CO2} emissions are expected to fall by 50\%. The demand for electricity would double, but the processes would be 16\% more energy-efficient \cite{ISTR2020}.

In order to achieve the estimated iron output needed to supply future demand, while still reducing the total \ce{CO2} emissions as a step towards carbon-neutral production, the iron and steel industry must
implement technologies that are pre-commercial today, along with others that are quickly gathering pace, as well as strategies to increase the efficiency of established technologies \cite{ISTR2020, yin2020strength}.

In the medium term, \ce{CO2} intensity can be lowered in two ways. Firstly, direct emissions can be mitigated by replacing coal in blast furnaces with biomass or hydrogen—or, in an intermediate step, with natural gas. Secondly, a drastic reduction in indirect emissions can be achieved by decarbonizing the heat generation, while applying renewable heat sources and electrification \cite{ISTR2020,QUADER2015594,madeddu2020}. It is expected that in a sustainable development context, emissions in the electricity sector can be reduced by up to 95\% \cite{ISTR2020}.

The direct reduction of iron (DRI) using hydrogen as an agent is suggested as one of the key pathways to reaching carbon neutrality in the sector \cite{ISTR2020,QUADER2015594,thefutureofhydrogen}. It adapts a technology that is already mature, operating in a configuration capable of delivering zero direct emissions of \ce{CO2}. With hydrogen produced by electrolysis, using renewable energies as sources, indirect emissions can be further mitigated. This concept is already being applied in a pilot plant in Sweden, which predicts an increase in output to as much as 1 Mt of iron per year in 2025 \cite{HYBRIT}. Other projects are being pursued by ArcelorMittal, which plans to start producing iron through hydrogen DRI in its plant in Hamburg in 2030 \cite{ArcelorMittal}, and ThyssenKrupp, which is due to commission its first commercial large-scale hydrogen DRI plant in 2024 \cite{thyssenkrupp} as part of its project to become carbon neutral in 2050.

Research is also being carried out on other innovative technologies such as electrochemical reduction by the direct electrolysis of iron oxides. These concepts involve the removal of oxygen from the raw material by passing an electric current through it, either in an aqueous solution \cite{siderwin} or through high-temperature (over 2000 $^\circ$C) molten oxide electrolytes \cite{wiencke2018,bostonmetal}. Its global reaction is the inverse of reaction (1), enhanced by catalysts. The advantage would be to avoid intermediate processes that inherently lead to energy losses. Both concepts, however, are still at the prototype stage, therefore it is not yet possible to predict when—or if—they will be commercially available.

For the purposes of the present work, a route is evaluated in which the direct reduction through hydrogen is the main source of primary iron as an energy carrier, considering its feasibility in a broader context of a green hydrogen economy \cite{thefutureofhydrogen}, as well as its potential in recycling iron oxides. With policies aimed at phasing out coal and other fossil fuels, along with massive investments in hydrogen electrolyzer facilities powered by renewable energy \cite{h2europe}, as well as the push for full decarbonization by the steel industry, it is expected that direct iron reduction will have a major role in achieving carbon neutrality in the iron and steel sector \cite{ISTR2020}.

\subsection{Scenario for the present and future of hydrogen}
\label{sec:H2_scenario}
Hydrogen is presently a very important intermediate in a number of processes. Around 73 Mt of pure \ce{H2} are produced annually, of which 38 Mt are used in the refining of petrol, and 31 Mt in the production of ammonia \cite{thefutureofhydrogen}. Only around 4 Mt are used in other applications, including transportation and heat generation \cite{thefutureofhydrogen}.

Production processes can be classified into three main categories: the first, the production of fossil-fuel-based hydrogen, involves the use of coal (black hydrogen), lignite (brown hydrogen) or natural gas (gray hydrogen), resulting in large amounts of \ce{CO2} emissions; the second, that of blue hydrogen, uses the same processes as with fossil-fuel-based hydrogen, but applying carbon capture, use and storage (CCUS) techniques to stop the \ce{CO2} from being released into the atmosphere; and the third, the production of green hydrogen, uses renewable energy sources, generating significantly lower emissions \cite{thefutureofhydrogen}.

Gray hydrogen is presently the most common route, accounting for around 75\% of the total yearly supply. In this process, \ce{H2} is mainly produced from the steam reforming of methane, producing syngas that is later separated, emitting \ce{CO2}. Due to the availability of coal in China, black hydrogen is the main route in that country, generated by gasification processes, resulting in syngas and accounting for 23\% of the global output. Green hydrogen and other pathways only account for 2\% of production, and only 0.1\% is produced from electrolysis \cite{thefutureofhydrogen}. When electricity comes from solar or wind farms, indirect emissions are also very low, having a negligible impact on global warming.

\subsubsection{Roadmap for green \ce{H2} production}
\label{sec:Roadmap_green_H2}
Clean hydrogen is gathering unprecedented momentum, and is presently one of the main focuses of energy policies and research worldwide. Due to its versatility, it is tipped as a key means of decarbonization, either as an energy carrier or as a chemical agent, in the hydrogen economy \cite{thefutureofhydrogen}. In this context, \ce{H2} is considered as a mean of storing and distributing energy from renewable sources, mitigating the issues of intermittency and enabling harnessed energy to be used for various purposes. For iron production, as previously mentioned, hydrogen can work in both roles, providing energy for the production processes and acting as a reduction agent for oxides.

Governmental and industrial strategies are already pointing in this direction. Germany, Italy and Greece have each announced plans to build 5 GW of electrolyzers by 2030. Spain and the Netherlands have plans for 4 GW each, and France 6.5 GW. These efforts are part of the European roadmap for hydrogen, which outlines the implementation of a total of 40 GW of electrolyzer capacity  by 2030 \cite{h2europe}. The German policy is notable, with the German Advisory Council on the Environment particularly acknowledging the capabilities of green hydrogen as an energy carrier, stabilizing the electricity grid and enabling energy trading between countries—which is especially important, considering that Germany is still expected to rely on energy imports for years to come \cite{SRU2021}. It is expected that, by 2045, 647 TWh of energy consumed in Germany will come from energy carriers, with a share of 458 TWh of green \ce{H2}. Only 9.3\% of these carriers will be produced domestically.

Despite its present-day costs, electrolysis is seen as playing a major role in future hydrogen production \cite{h2europe}. In places where solar and wind electricity are low-cost, such as in Northern African countries, in the Middle East and in Spain, it may soon become commercially interesting to produce hydrogen from renewable sources, with prices comparable to or lower than those of gray hydrogen \cite{IRENA2020}. In the Arabic Peninsula, for example, photovoltaic (PV) solar power has already reached values as low as USD 0.0104/kWh—more than five times lower than the cheapest coal production anywhere in the world, estimated at USD 0.055/kWh \cite{IRENA2020}. The recent crisis in Ukraine has led to price fluctuations and increased the risks of disruption in the fossil-fuel supply chain, promoting a faster transition to renewables in Europe. Altogether, these facts are pointing to larger scales and lower costs for green hydrogen production, which will be soon widely available for a fraction of the present-day price \cite{IRENA2020}.

Two of the main challenges of hydrogen are storage and transportation. Systems to store and deliver the substance in gaseous form require safety measures to avoid explosion and flashback risks, as well as losses to the environment \cite{thefutureofhydrogen,BERSTAD2022111772}. Moreover, considering the lack of delivery routes (such as pipelines for natural gas) between the location of the expected producers and consumers, hydrogen delivery pipelines would have to be built from scratch, with different requirements. In this respect, there is a strong synergy with the use of iron as an energy carrier as an auxiliary step to deliver energy from hydrogen \cite{BERGTHORSON2015368}. Iron is stable and can easily be stored and transported similarly to other bulk goods, such as coal and iron ore, making use of existing shipping routes and equipment designed for these substances. This is of utmost importance before hydrogen costs are lowered globally, production is scaled up and the transportation and storage of gaseous hydrogen is solved. 

In the present work, the electricity requirements include the amount used to produce hydrogen for the reduction part of the cycle, as well as the heat required for the reduction processes. This extra energy requirement is taken into consideration when evaluating the global efficiency of the reduction process. The heat can also provided directly by electricity, but the process could be impacted by intermittency on renewable power generation. In the present work projections are made using the alternative route, which avoids the intermittency problem, as it runs completely on \ce{H2} that would be already available as a reducing agent. 

\section{Estimation of round-trip efficiencies}

A brief comparison of round-trip efficiencies (power-to-power) is proposed for a hypothetical case in which renewable electricity production, \ce{H2} generation and iron reduction takes place in North Africa, and the materials are transported via bulk carrier ships to Rotterdam. Conversion takes place in a retrofitted coal power plant and iron oxides are transported back to North Africa. The numbers in brackets indicate the energetic efficiency for each process step:

\begin{itemize}
    \item Iron (Electrochemical route—EC): iron reduction EC (80\%), iron powder generation including melting (87\%), transport by ship from Casablanca to Rotterdam (outward/return) (93--96\%) \cite{DIRVEN201852}, steam power plant (50\%). Round-trip efficiency 32--33 \%.
    \item Iron (Thermochemical route—TC): \ce{H2} production by electrolysis (70--80\%), iron reduction-TC (80\%), transport by ship from Casablanca to Rotterdam (outward/return) (93--96\%) \cite{DIRVEN201852}, steam power plant (50\%). Round-trip efficiency 26--31 \%.
    \item \ce{H2}: \ce{H2} production by electrolysis (70--80\%), liquid \ce{H2} transport by ship from Casablanca to Rotterdam (55-80\%) \cite{IRENA2020}, combined cycle gas turbine process (60\%). Round-trip efficiency 23--38\%.
\end{itemize}

While most of the reported single-step efficiencies are well-known, the transport energy efficiency is still an unresolved topic. Readers are referred to the IRENA study \cite{IRENA2020} for the estimations of energy loss for long-distance transportation of liquid \ce{H2} by ship and the recent study by Dirven et al. \cite{DIRVEN201852} on energy loss during the transportation of iron. It is clear that, when \ce{H2} is produced locally, its direct use can be more efficient. This also requires the local availability of cheap, abundant renewable electricity. Efficiencies are significantly lower when the transportation and storage of \ce{H2} is necessary. In such conditions, the iron-fuel cycle becomes more efficient. It is also possible to use the extensive infrastructure already available to transport the iron fuel and to generate power. A complete and detailed assessment of the round-trip process efficiencies and a comparison of power-to-power technologies using different energy carriers in a variety of scenarios are recommended for future investigations.

\section{Projections for replacing coal with iron}
\label{sec:Projections}

In this section, estimations and projections are given for replacing coal with iron. First, general correlations are proposed. They require the annual electricity output from coal power plants as an input to assess the materials and infrastructure required. The correlations are then applied to estimate the requirements in several retrofitting scales.

\subsection{Correlations for the estimations}
\begin{figure}[h]
	\begin{center}
	\includegraphics[width=1\textwidth]{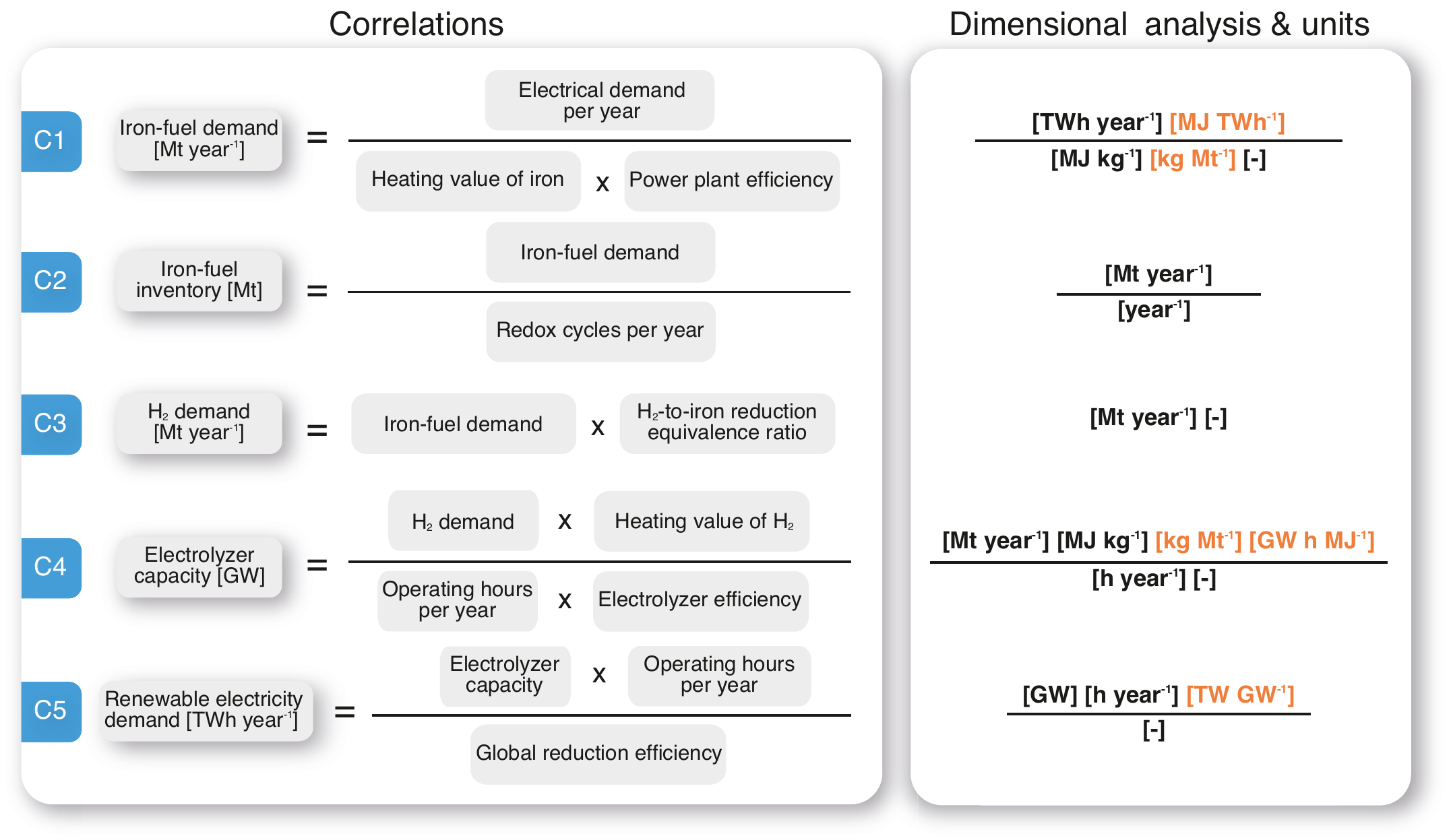}
	\caption{\textbf{Correlations for estimating the material and infrastructure required to replace coal with iron.}}
    \label{fig:equations}
	\end{center}
\end{figure}

In this proposed concept, replacing coal would require the availability of not only iron and hydrogen as raw materials, but also the infrastructure to generate the renewable electricity needed to produce the \ce{H2} and to reduce the iron. Building the necessary iron inventory requires planning on a large scale, which is directly proportional to the frequency of redox cycles performed in a certain period of time. Moreover, this frequency is also directly related to the distance and transportation time between the power plants and the reduction sites. In order to facilitate the estimation of these values for different scales and scenarios, a set of general correlations was developed (Fig.~\ref{fig:equations}).
These equations take into account the quantification of the most essential materials and infrastructure: the metal-fuel iron, the reducing agent \ce{H2} and the electrolyzer capacity for its production, and the renewable electricity to power the electrolyzers. The correlations take into consideration essential well-known variables (e.g., properties of iron and hydrogen) and case-specific estimated variables (i.e., efficiencies, operating hours, frequency of redox cycles).

Individual correlations are proposed for (C1) the iron demand yearly, (C2) the size of the iron inventory, (C3) the \ce{H2} demand for reduction, (C4) the required electrolyzer capacity and (C5) the renewable electricity demand. The correlations are proposed employing typical units for each variable. A dimensional analysis is reported on the right-hand side of the correlations, while the necessary unit conversions are highlighted in orange. A detailed explanation on the development of the correlations is available in the appendix. The constants and units used for the projections in the present work are reported in Table \ref{table:constants}.

For the estimations in the present work, case-specific constants were used, see Table \ref{table:constants}. A detailed explanation of the numbers adopted is presented in the appendix.

\begin{table}[h]\small
\caption{Constants used in the projections of material requirements for the retrofitting framework.}
\centering
\begin{tabular}{l c c}
    \hline
    Correlation constant & Values & Units \\
    \hline
    LHV of iron    & 7.36 & [MJ kg$^{-1}$]   \\
    LHV of \ce{H2}  & 119.96 & [MJ kg$^{-1}$] \\
    \ce{H2}-to-iron reduction equivalence ratio  & 0.0534 & [Mt Mt$^{-1}$]\\
    Power plant efficiency  & 40 & [\%] \\
    Electrolyzer efficiency   & 70 & [\%]   \\
    Global reduction efficiency    & 85 & [\%]  \\
    Redox cycles per year  & 1, 2, 4, 12 & [year$^{-1}$] \\
    Operating hours per year   & 2000, 4000 and 8000&  [h year$^{-1}$]\\
    \hline
\end{tabular}
\label{table:constants}
\end{table}

\subsection{Estimation of iron requirements for retrofitting coal power plants}

In order to estimate the figures for replacing coal, the demand for coal-generated electricity must be considered. Table \ref{table:coal_elec} reports the share of electricity from coal worldwide. China's electricity output from coal increased continuously in the 2016--2020 period. India exhibited the same pattern as the world as a whole: the historic peak was reached in 2018, followed by two years of declining generation. Coal electricity figures decreased in Germany, in the EU28 and in the United States in the same period.
The International Energy Agency (IEA) forecasts that a new peak in global coal-fired power generation will be reached in 2021 due to the surge in demand following the pandemic, with this increase occurring in all the countries listed~\cite{Coal2021}. However, of these countries, new historical peaks would only be found in India and China.
Therefore, the 2018 electrical output was considerd to be more appropriate for the scientific analyses and projections within the study, as it avoids the short-term effects caused by the pandemic.

\begin{table}[h]\small
\caption{Share of electricity from coal by country/region, 2016-2021}
\centering
\begin{tabular}{l c c c c c c}
    \hline
    Country/Region & \multicolumn{6}{c}{Absolute Share [TWh]} \\
     & 2016 & 2017 & 2018 & 2019 & 2020 & 2021* \cite{Coal2021} \\
    \hline
    Germany \cite{emberEurope} & 262 & 241 & 228 & 171 & 134 & 161\\
    EU+UK \cite{IEAwebsite} & 737 & 710 & 660 & 498 & 393 & 472  \\
    Europe \cite{IEAwebsite} & 965 & 928 & 891 & 726 & n.a. &  n.a. \\
    USA \cite{emberEurope}& 1239 & 1206 & 1146 & 965 & 774 & 929 \\
    India \cite{emberEurope}& 936 & 973 & 1025 & 999 & 947 & 1061 \\
    China \cite{emberEurope}& 3946 & 4178 & 4483 & 4554 & 4631 & 5048 \\
    Worldwide \cite{BP2021}& 9451 & 9806 & 10091 & 9826 & 9421 & 10350\\
    \hline
    \end{tabular}
\label{table:coal_elec}
\end{table}

\begin{figure}[btbp]
		\begin{center}
        \includegraphics[width=0.9\textwidth]{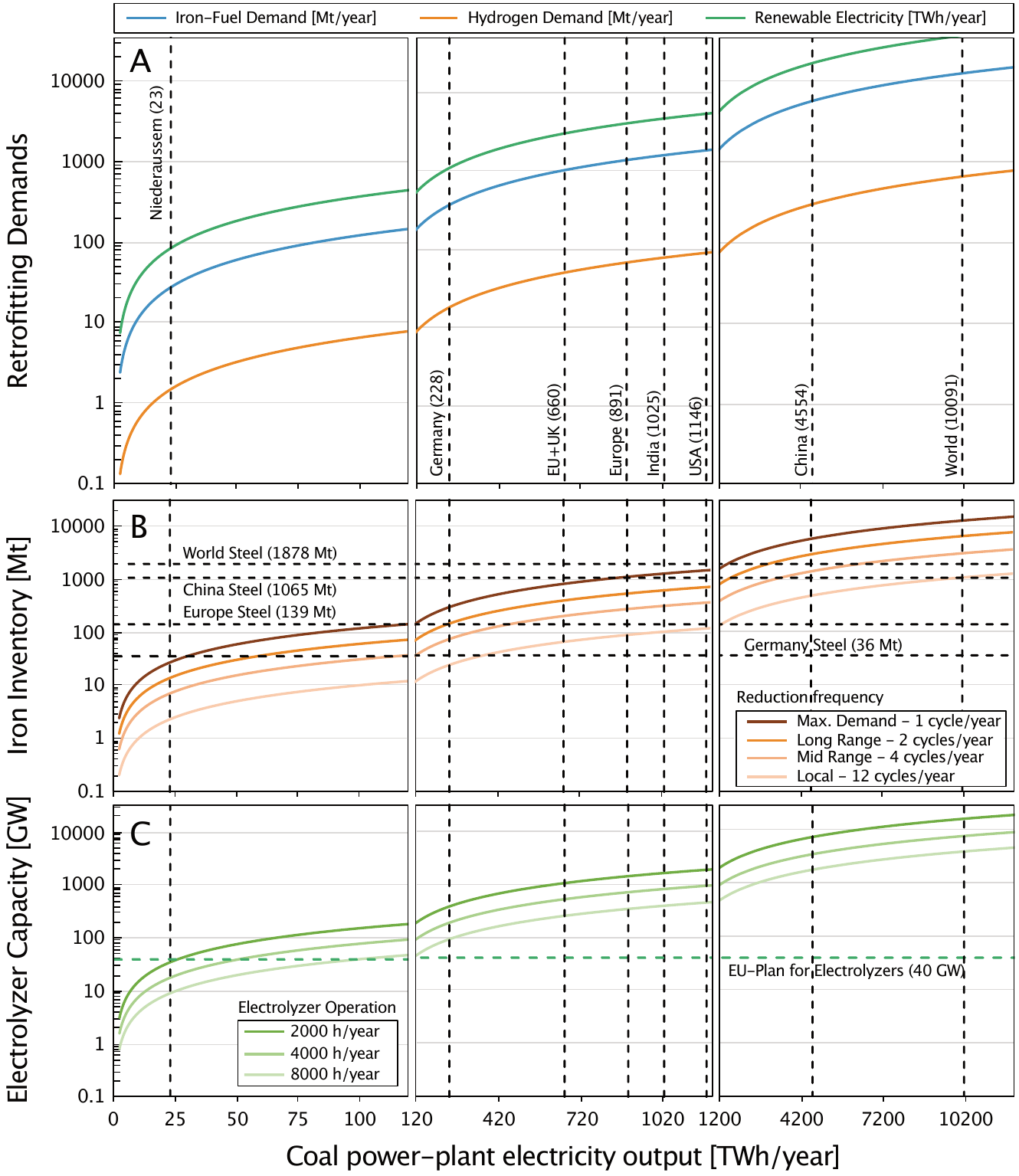}
        \end{center}
        \vspace{-10pt}
        \caption{\textbf{Projections of demands for retrofitting coal power plants.} Vertical lines report the electricity produced using coal in the Niederaussem power station, in several countries and worldwide \cite{Coal2021,WEO2020,BP2021} in 2018. Top row (A): Demands of iron, \ce{H2} and renewable electricity in parity with the coal-fired electricity produced (correlations C1, C3 and C5). Middle row (B): Necessary iron inventory as a function of the number of reduction-oxidation cycles per year (correlation C2). Horizontal lines report yearly steel productions. Bottom row (C): Necessary electrolyzer capacity as a function of operating hours per year (correlation C4).}
		\label{fig:projections}
\end{figure}

Fig. \ref{fig:projections} reports a multi-scale projection of iron, hydrogen and renewable electricity demands, obtained using the correlations reported in the section. For the analysis proposed in the present work, first a single power station in Germany is considered, then it is compared with two larger scales: European (including the European Union and the United Kingdom) and worldwide. Any other scales can be analyzed using the same method, taking into account the efficiencies and characteristics of the region to be analysed.

The example taken is the Niederaussem coal power station (Germany), consisting of seven units, with a capacity of 3641 MW \cite{niederaussem}. In 2018, the station was among the ten most carbon-emitting plants worldwide with 27.2 Mt of \ce{CO2} and having an emission intensity of 1.2 kg of \ce{CO2} per kWh—45\% higher than the average fossil-fuel index for Germany \cite{Grant_2021}—generating around 23 TWh of electricity.

It can therefore be estimated that to cover 100\% of Niederaussem's electricity production, using correlation C1, 28 Mt of iron need to be burned per year. However, the total iron inventory required depends on the distance to the reduction facilities where iron is regenerated. The number of redox cycles that can be performed per year depends inversely on the transportation time. Assuming that iron reduction is local, the iron could be regenerated monthly, shrinking the inventory requirement by 12 times. At mid-range distances (e.g., Germany--Morocco or Germany--Southern Spain), the two-way transportation time reduces the number of possible redox cycles per year. At long-range distances (e.g., Germany--Canada or Germany--Australia), this number decreases even further. Therefore, those 28 Mt of iron required per year represent 7 and 2.3 Mt of inventory in the case of four and twelve cycles per year, respectively, using correlation C2. Similar estimations for the other scales can be obtained from Fig. \ref{fig:projections}B, showing the overall required iron inventory as a function of the number of redox cycles.

\subsubsection{Ramping up production for building iron inventories}
Securing iron fuel inventories requires the availability of iron ore in the market and the capacity to produce reduced iron from it. The steel industry has the highest capacity to reduce iron. The correct scaling of production to meet the increased demand for both iron ore and reduced iron is therefore essential. The projections for the iron fuel demand can be compared with the corresponding steel production figures for the same regions, for a direct assessment of how much more local iron reduction would be necessary to build a certain iron inventory, and an indirect assessment of how much more iron ore would need to be secured from the markets. Reported as horizontal dashed lines in Fig. \ref{fig:projections}B, crude steel production in 2020 was 35.7, 139.2 and 1878 Mt in Germany, the European Union (including the United Kingdom), and the world, respectively \cite{worldsteelinfigures}. 

Returning to the case of Niederaussem power plant, assuming that it would be possible to complete four redox cycles per year, the plant needs an iron inventory of 7 Mt. With respect to the German crude steel production of 35.7 Mt, these 7 Mt of iron inventory represent 19.6\% of one year's production output. Although that is a large fraction, it could certainly be secured by building up the inventory over a few years. Distributed over a period of 5 years, that percentage drops near 4\% of the production output. In other words, ramping up the domestic iron reduction capacity by 4\% and maintaining the production level for a 5-year period would meet the increased demand for iron to compile the Niederaussem power plant metal-fuel inventory.

However, for large-scale retrofitting, these iron inventories need to be built up over longer periods. Similarly to the case of Niederaussem, the iron inventories required to fully retrofit the power plants in Germany, Europe and the world are compared with the corresponding yearly steel production figures in each region. The necessary production ramp-ups were calculated considering scenarios of 10 and 20 years to build up those inventories. For Germany, ramping up the domestic iron reduction capacity by 9.7\% and maintaining that level is enough to compile the necessary inventory over a period of 20 years. This percentage increases to 19.4\% for a 10-year period. The total German inventory represents nearly two years of accumulated national steel production (194\%). In the case of Europe, ramping up its local iron reduction capacity by 7.2\% and maintaining it at that level meets the 20-year planning requirements. On a worldwide scale, a 20-year plan would require that capacity to be ramped up by 8.2\% and maintained at that level.

The steel industry predicts that by 2050, in a Stated Policies Scenario, the global yearly demand for iron and steel will reach 2.5 Gt \cite{ISTR2020}. This corresponds to a forecast 30\% increase in demand compared to the current 1.9 Gt. The increase in demand from the energy sector can likely be absorbed by this market, especially if there is a gradual yearly production increase accompanied by national incentive plans. Long-term plans for building up inventories also include the gradual construction of the green-iron reduction infrastructure; both are necessary for implementing a retrofitting plan.

\subsection{Estimation of hydrogen requirements for iron reduction}
The estimations for the Niederaussem power plant (23~TWh) reveal that a considerable amount of \ce{H2} is required (1.5~Mt per year, calculated using correlation C3), as well as an infrastructure comprised of electrolyzers (17.9~GW, using correlation C4) and renewable electricity generation (71.6~TWh, using correlation C5). Considering that in 2020, Germany produced 131~TWh of electricity from wind alone \cite{WEO2020}, producing the 71.6~TWh needed to retrofit this large power plant is a feasible task for the upcoming decades, when the infrastructure, technology and share of renewables will be considerably improved.
Expanding the retrofitting plans to national and regional scales would create a new level of demand. This increased demand can only be met if the infrastructure for the reduction part of the cycle is planned and built in coordination with the retrofitting process. Fig.~\ref{fig:projections} shows estimations for the complete retrofitting of the European coal power plants can be observed, pointing to a yearly requirement of 803~Mt of iron, 2.4~PWh of renewable energy to produce 43~Mt of \ce{H2}, employing 514~GW of electrolyzers operating 4000 hours a year. Taking 4 redox cycles a year, the required iron inventory amounts to 200~Mt. In a 20-year plan, 10 additional Mt per year would be added to the iron inventory, while the infrastructure construction would need to cover for an additional 26~GW of electrolyzer capacity and the production of an additional 120~TWh of clean electricity per year, producing an additional 2.2~Mt of \ce{H2} annually. A more ambitious 10-year plan would require these yearly quantities to be doubled.

Examining the current European and German plans, the required infrastructure for this solution is ambitious. However, this increased demand could promote the maturation of electrolysis technology, bringing efficiencies up and costs down as the implementation plans are boosted. In the long term, clean electricity will be abundant, meaning that iron fuel redox cycles would be very attractive as a way of storing energy in periods with abundant sun hours and wind to make up for periods of scarcity, as well as providing the infrastructure for carbon-neutral steel production.

\section{Conclusions}
\label{sec:Conclusions}
Climate change, mainly caused by GHG emissions, is a topic of increasing global concern. The current situation was firmly assessed during the COP26 summit, highlighting the fact that coal combustion is one of the largest contributors to \ce{CO2} emissions. Phasing down coal—except when carbon capture, storage and/or use processes exist—was agreed unanimously as an urgent measure to reach the climate goals established in the Paris Agreement. Moreover, several countries have committed to phasing coal out as soon as possible. For many nations, this transition will be a great challenge, requiring adequate strategies and planning. Germany previously announced a national phase-out by 2038 at the latest, and the new government has announced plans to reach that goal in 2030.

General correlations were proposed for estimating the necessary resources and infrastructure required to retrofit coal power plants for iron combustion. The formulation used is not case-specific and can easily be extrapolated for other contexts and implementation scenarios and scales. 

Using this formulation, three different retrofitting scales were evaluated: first a single power station, followed by the European context, then worldwide. This study indicated that Germany's high-capacity Niederaussem coal power station (3641~MW), which generates about 23~TWh of electricity annually, would need to burn 28~Mt of iron. That corresponds to an iron inventory of 2.3~Mt of iron, if reduction is done locally, requiring 1.5~Mt of \ce{H2}, which could be produced by 17.9~GW of electrolyzer capacity (operating 4000 hours a year). The quantification was extended to the European scale, resulting in iron requirements of 803 Mt per year to generate the equivalent of 660~TWh of electricity. The total necessary inventory of this energy carrier was estimated as a function of the number of redox cycles per year. Performing the reduction part of the cycle would require 43~Mt of \ce{H2} a year, which could be produced using 514~GW of electrolyzer capacity, powered by approximately 2.4~PWh of clean energy.  The building up of the necessary iron inventories and infrastructure was evaluated for 10- and 20-year implementation periods, revealing that reasonable ramping-up levels of iron reduction capacity would meet these long-term implementation plans. 

The results of this study reveal a great opportunity to overcome the challenges of phasing out coal, supporting the energy transition plans. The proposed solution would not only preserve and take advantage of existing infrastructure (coal thermal power plants) and create an asset of energy carriers (iron reserves) that increase energy security, but also promote the development and implementation of the technology and infrastructure needed for renewable energy, hydrogen and green steel production. Round-trip comparisons revealed that this concept results in promising efficiencies, in particular for regions where renewable energy is not abundant.

Accomplishing a zero-emission economy transition will be a difficult task for many countries, but an ambitious large-scale plan of this kind becomes feasible when it is implemented gradually, building up the necessary infrastructure in parallel. The challenge of establishing large-scale \ce{H2}-reduction systems for iron can be tackled in tandem with the steel industry, which is moving towards climate neutrality by 2050. The results presented in this work can {further} be used as a model for planning the retrofitting of power plants that have already been deactivated, in countries where coal is already being phased out, creating a back-up source of power generation. They can also provide support to other coal-dependent countries worldwide and in Europe. A thorough evaluation of the technological and economic feasibility must be addressed by the scientific community, possibly promoting public opinion in favor of this strategy and affecting policymakers' decisions.

\section*{Acknowledgments}
This work was funded by the Hessian Ministry of Higher Education, Research, Science and the Arts—Clean Circles cluster project.
\makeatletter
\renewcommand{\thesection}{A}
\renewcommand{\thefigure}{A\@arabic\c@figure}
\renewcommand{\thetable}{A\@arabic\c@table}
\renewcommand{\theequation}{A\@arabic\c@equation}

\setcounter{figure}{0}
\setcounter{table}{0}
\setcounter{equation}{0}

\makeatother

\section*{Appendix: }

\makeatletter 
\renewcommand{\thesection}{A\c@section}
\renewcommand{\thefigure}{A\@arabic\c@figure}
\renewcommand{\thetable}{A\@arabic\c@table}
\renewcommand{\theequation}{A\@arabic\c@equation}

\setcounter{section}{1}
\setcounter{figure}{0}
\setcounter{table}{0}
\setcounter{equation}{0}

\makeatother

\section*{Appendix A: Materials required to replace coal with iron}
In a typical modern coal power plant, conversion from primary to electrical energy reaches an efficiency range between 38\% and 42\% \cite{sargentandlundy2009}. The value of 40\% is chosen for these calculations. Taking the highest-rank bituminous coals with a LHV of 36 MJ kg$^{-1}$ and a density of 1200 kg m$^{-3}$, 0.25 kg of coal are necessary to produce 1 kWh of electrical energy.
When retrofitting a power plant of this kind for iron combustion, a similar conversion efficiency can be assumed. Since the technology is not currently in use in commercial units, a similar efficiency to that of coal can be seen as a lower boundary, especially in view of irreversibilities such as the presence of impurities in coal and the possible partial conversion of iron \cite{bergthorson2018}. This value leads to 1.216 kg of iron being required per kWh produced. This results in 4.86 times more mass, but only 75 \% the volume of iron in comparison to coal.
\begin{center}
0.25 kg of bit. coal $\rightarrow$ 1 kWh electrical energy $\leftarrow$ 1.216 kg of iron\\
1 kg of bit. coal $\leftrightarrow$ 4.86 kg of iron\\
1 m$^{3}$ of bit. coal $\leftrightarrow$ 0.75 m$^{3}$ of iron\\
\end{center}

While iron combustion produces only non-toxic solid oxides, coal combustion produces very high levels of \ce{CO2} per energy unit. For simplicity, coal is assumed to be pure carbon here. Therefore, the direct emissions from coal combustion can be roughly estimated based on the complete conversion of the carbon present in the coal into \ce{CO2}:

\begin{equation}
   Coal(C) + \ce{O2} \Leftrightarrow \ce{CO2}
\end{equation}

\begin{center}
12 g + 32 g $\rightarrow$ 44 g
\end{center}

For simplicity, from each carbon atom in coal, one molecule of \ce{CO2} is emitted, which results in a mass ratio of 3.67 times.

\begin{center}
250 g of bit. coal$\rightarrow$ 917 g of \ce{CO2} $\rightarrow$ 1 kWh electrical energy \\
\end{center}

The projections for the reduction of oxides start out from the stoichiometry of the process. Considering that the main product of iron combustion is \ce{Fe2O3}, the global reduction process is:

\begin{equation}
   Fe_2O_3 + 3 \cdot \ce{H2} \Leftrightarrow 2\cdot \ce{Fe} + 3 \cdot \ce{H2O}
\end{equation}

From this global reaction, it is seen that:
\begin{center}
3 mols of \ce{H2} $\rightarrow$ 2 mols of iron\\
In terms of mass:\\
6 kg of \ce{H2} $\rightarrow$ 111.68 kg of iron\\
Normalizing these quantities results in:\\
1 kg of \ce{H2} $\rightarrow$ 18.61 kg of iron\\
53.4 g of \ce{H2} $\rightarrow$ 1 kg of iron\\
\end{center}

In order to produce one unit of mass of reduced iron, 0.0534 units of mass of hydrogen are required.

The key step when producing hydrogen from renewables involves the electrolysis of water by passing an electric current through it, breaking it into \ce{H2} and \ce{O2}, following the global reaction:

\begin{equation}
   \ce{H2O} + electricity \rightarrow \ce{H2} + \frac{1}{2} \cdot \ce{O2}
   \label{eq:electrolysis}
\end{equation}

The process is very energy-intensive, but generates zero direct \ce{CO2} emissions, while exhibiting efficiencies between 60\% and 81\% (conversion of electrical energy into heat value of hydrogen) \cite{thefutureofhydrogen}. Electrolyzer technology is improving, with current estimations of nearly 80\% efficiency for commercial applications in the near future \cite{thefutureofhydrogen}. Taking a conservative efficiency of 70\%, and the heating value of \ce{H2} of 119.96 MJ/kg, it can be estimated that:

\begin{center}
1 kg \ce{H2} $\rightarrow$ 119.96 MJ (heat value)\\
Electrolyzer efficiency = 70\% \\
119.96 MJ / 0.7 = 171.4 MJ (electrical energy) $\rightarrow$ 1 kg \ce{H2}\\
Converting to kWh:\\
47.6 kWh electrical energy $\rightarrow$ 1 kg \ce{H2}\\
\end{center}

The production of one kg of \ce{H2} requires 46.8 KWh of electrical energy. From this equivalence, it is also possible to derive the energy required to produce the equivalent mass of reduced iron:

\begin{center}
Required electrical energy:\\
47.6 kWh electrical energy $\rightarrow$ 1 kg \ce{H2} $\rightarrow$ 18.61 kg of iron \\
Normalizing it to one unity of iron:\\
2.56 kWh \ce{H2} $\rightarrow$ 53.4 g \ce{H2} $\rightarrow$ 1 kg of iron \\
\end{center}

Electrolyzer operation would depend on the source of renewable energy. Is is fair to assume that an electrolyzer plant would be able to operate for an average of 4000 hours a year when powered by off-shore wind power \cite{WindMonitor}. Solar photo-voltaic and on-shore wind power could provide an average of 2000 hours a year \cite{WindMonitor,Fraunhofer_PVfacts,Fasihi2017}. A more constant supply could be obtained using a combination of wind, solar photo-voltaic and solar thermal, increasing the overall operating hours to a possible 8000 hours a year \cite{Fasihi2017}. The capacity of these plants is described by their electrical energy input. For a 1 GW plant operating 4000 hours a year, it is possible to derive the consumption of electricity and the expected production of \ce{H2}:

\begin{center}
Electricity consumption:\\
1 GW capacity x 4000 h/year $\rightarrow$ 4 TWh/year\\
Equivalence to \ce{H2} production:\\
4 TWh = 84.03 kt \ce{H2} \\
Normalizing to one unit of \ce{H2}:\\
1 Mt \ce{H2} = 11.9 GW electrolyzer capacity = 47.6 TWh electricity
\end{center}

The reduction of iron by \ce{H2} requires additional energy input, which must be taken into account when estimating the overall energy required by the cycles. Taking the chemical reactions, \ce{Fe} oxidation to \ce{Fe2O3} has a reaction enthalpy of -820 MJ/kmol. The enthalpy of \ce{Fe2O3} reduction to \ce{Fe} by \ce{H2} is +100 MJ/kmol, from which 720 MJ/kmol is obtained from the enthalpy of \ce{H2} itself. Therefore, an additional amount of electricity in the form of heat, amounting for 13.9 \% of the required energy to produce the \ce{H2}, must be taken into consideration. This contribution can be included directly in the estimation of required renewable energy, assuming an efficiency factor of the reduction process. For simplicity, a factor of 85 \% efficiency is assumed.

These estimates also make it possible to derive the \ce{H2} input, electrolyzer capacity, and electrical energy required to retrofit coal power plants. These numbers do not depend on the number of redox cycles.

\bibliography{literature.bib}
\bibliographystyle{unsrtnat}

\clearpage

\end{document}